 \documentclass[epj]{svjour}

 \usepackage{epsfig}

\newcommand\cpc[3]   {{Comput.\ Phys.\ Commun.\ }{\bf #1} (#2) #3}

\newcommand\epjc[3]  {{Eur.\ Phys.\ J. }{\bf C #1} (#2) #3}

\newcommand\jhep[3]  {{J. High Energy Phys.\ }{\bf #1} (#2) #3}

\newcommand\nc[3]    {{Nuovo Cim.\ }{\bf #1} (#2) #3}

\newcommand\plb[3]   {{Phys.\ Lett.\ }{\bf B #1} (#2) #3}
\newcommand\pr[3]    {{Phys.\ Rev.\ }{\bf #1} (#2) #3}

\newcommand\prd[3]   {{Phys.\ Rev.\ }{\bf D #1} (#2) #3}

\newcommand\prep[3]  {{Phys.\ Rept.\ }{\bf #1} (#2) #3}
\newcommand\prl[3]   {{Phys.\ Rev.\ Lett.\ }{\bf #1} (#2) #3}

\newcommand\sjnp[3]  {{Sov.\ J.\ Nucl.\ Phys.\ }{\bf #1} (#2) #3}

\newcommand\yf[3]    {{Yad.\ Fiz.\ }{\bf #1} (#2) #3}
\newcommand\zpc[3]   {{Z.\ Physik }{\bf C #1} (#2) #3}

\newcommand\ibid[3]{{ibid.\ }{\bf #1} (#2) #3}

\newcommand{\hepph}[1]{{hep-ph/#1}}

\newcommand{\hepex}[1]{{hep-ex/#1}}

 \title{Regge-cascade hadronization}

 \author{K.~Odagiri}

 \institute{
  Institute of Physics, Academia Sinica, Nankang, Taipei, Taiwan 11529,
  The Republic of China}

\begin{document}

 \abstract{
  We argue that the evolution of coloured partons into colour-singlet 
hadrons has approximate factorization into an extended parton-shower 
phase and a colour-singlet resonance--pole phase.
  The amplitude for the conversion of colour connected partons into 
hadrons necessarily resembles Regge-pole amplitudes since $q\bar q$ 
resonance amplitudes and Regge-pole amplitudes are related by duality. A 
`Regge-cascade' factorization property of the $N$-point Veneziano 
amplitude provides further justification of this protocol.
  This latter factorization property, in turn, allows the construction 
of general multi-hadron amplitudes in amplitude-squared factorized form 
from $(1\to2)$ link amplitudes.
  We suggest an algorithm with cascade-decay configuration, ordered in 
the transverse momentum, suitable for Monte-Carlo simulation.
  We make a simple implementation of this procedure in Herwig++, 
obtaining some improvement to the description of the event-shape 
distributions at LEP.
 }

 \date{June 27, 2006}

 \maketitle

 \section{Introduction}\label{sec_introduction}

  By `hadronization', we loosely mean the final phase in the process of 
jet formation, where coloured partons at the end of the parton shower 
turn into colour-singlet hadrons.

  The hadronization phase is presumably not completely separable from 
the perturbative phase, but approximations can be made, and we can for 
instance set the perturbative coupling to zero in the hadronization 
phase and vice versa. This is, in effect, the approach ordinarily taken 
in the Monte-Carlo event generators \cite{herwig,pythia}.

  A better approximation may be to allow the perturbative coupling to 
extend into the hadronization phase, and continue the perturbative 
evolution down to zero \cite{odagirihadronization}.

  The universality, i.e., that it does not depend on, for instance, 
where the other partons are, of the cut-off in the first case and the 
coupling in the second case can be justified in the Gribov confinement 
picture \cite{gribov}. Although the complete treatment of the gluon 
Green's function is lacking at present, for the quark Green's function, 
Gribov's method indicates that its behaviour is governed by a universal 
equation containing both the gluonic semi-perturbative and the 
long-distance super-critical contributions.
  We can then separate out the semi-perturbative contribution by means 
of the effective coupling procedure, so that the remaining dynamics, of 
hadronization, would be a predominantly colour-singlet interaction, 
mediated by resonances and poles. In other words, the coloured partons 
that remain at the end of the extended parton-shower phase turn into 
hadrons by an interaction which is effectively colour-singlet.

  Colour preconfinement \cite{amati_veneziano,preconfinement} dictates 
that at the end of the parton shower, the colour-connected parton pairs, 
i.e., the colour dipoles, have a mass spectrum with a characteristic 
scale of a few times the parton-shower cut-off, and this mass scale 
normally comes out to be about 1~GeV. However, this is violated in 
low-$p_T$ jets \cite{odagirisue}.

  As is the case in the Monte Carlo event generators, let us consider 
these colour-preconfined units as the starting point of hadronization. 
The colour-connected partons then exchange objects that are effectively 
colour-singlet to turn into hadrons.

  A justification for the resemblance of the confining dynamics with 
colour-singlet exchange is in duality, i.e., the observation that the 
summation over resonance states reproduces the dynamical behaviour 
characteristic of Regge poles, and vice versa \cite{collins,ddln}.

 \begin{figure}[ht]{
 \centerline{\epsfig{file=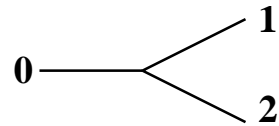, width=3.5cm}}
 \caption{A link in the cascade-decay chain.
 \label{fig_twobodydecay}}}\end{figure}

  In this paper, we concretize this statement by observing that the 
explicitly dual $N$-point Veneziano (i.e., the open bosonic string) 
amplitude \cite{veneziano,koba-nielsen} satisfies an amplitude squared 
factorization property that corresponds to cascade decay, where each 
vertex, shown in fig.~\ref{fig_twobodydecay}, has Regge-like angular 
behaviour, except in the final decay.

  Long-distance dynamics can then be treated as a series of Regge-like 
decay. This is literally true when the decaying unit is heavy, but 
because of semi-local duality \cite{finite_energy_sumrule}, Regge theory 
often remains a good approximation even in the low-energy region, or the 
low-mass region of the colour-preconfined units, where it is not 
formally supposed to be applicable, particularly after the resonances 
and dips have been integrated over.

  Combining this result with the above statement of approximate 
factorization of semi-perturbative and hadronization phases, we have a 
picture of hadronization where partons evolve via the parton shower and 
hadrons evolve via the Regge cascade.
  The exchange in the hadronization phase normally consists of the meson 
trajectories, but the baryons and even the pomeron trajectories are also 
allowed.

  The formalism is developed between secs.~\ref{sec_regge_two_to_two} 
and \ref{sec_fragmentation}.

  There is ambiguity as to whether, and to what extent, the forced $g\to 
q\bar q$ splitting at the end of the parton shower, which is adopted by 
HERWIG \cite{herwig} more or less for the sake of convenience, occurs in 
reality. In ref.~\cite{odagirihadronization}, we argued that there may 
be physical origin to the low-energy enhancement of the $g\to q\bar q$ 
splitting, due to the string tension `pulling apart' the colour octet. 
Another possibility would be that the gluon acquires a pole mass and 
hence can decay into two light quarks over a finite time. Even in this 
case, some of the gluons may remain intact, as all gluons do in PYTHIA 
\cite{pythia}. Since finite-time effects (gluon decay) are not 
completely separable from infinite-time effects (colour-singlet 
interaction), the latter effect cannot be neglected.
  In sec.~\ref{sec_montecarlo_threebody}, we discuss the hadronic 
observables that are potentially sensitive to this aspect of 
hadronization.

  In sec.~\ref{sec_implementation}, we examine the implementation of a 
simpified procedure based on the preceding discussions in Herwig++. We 
compute a number of jet observables at the $Z^0$ pole and compare 
against the Herwig++ and LEP numbers.

 \section{The $(2\to2)$ Regge amplitude}\label{sec_regge_two_to_two}

  The amplitude for $(2\to2)$ scattering with Mandelstam variables $s$ 
and $t$, omitting the signature factor, is:
 \begin{equation}
  A=\beta(t)\Gamma\left(\ell-\alpha(t)\right)(-s/s_0)^{\alpha(t)}.
  \label{eqn_regge_two_to_two}
 \end{equation}
  $\beta(t)$ is the coupling factor. Either $\beta(t)$ or 
$\beta(t)\Gamma\left(\ell-\alpha(t)\right)$ is often taken to be 
constant.
  $\alpha(t)$ is the $t$-channel trajectory and $\ell$ is the spin of 
its lowest-lying member. $s_0$ is the Regge characteristic scale. We 
identify $s$ with the dipole mass squared later on.

  The leading flavour-singlet trajectory is the the near-degenerate 
$\rho/\omega/f/a$ family:
 \begin{equation}
  \alpha(t)=\alpha_0+\alpha' t\approx 0.5+0.9t, \qquad \ell=1.
  \label{eqn_traj_roomaf}
 \end{equation}
  $t$ is measured in GeV$^2$. When $\sqrt{s}$ becomes large, above 
10~GeV or so, we should also consider the pomeron contribution.
  The typical transverse momentum generated by hadronization, in units 
of GeV, is then:
 \begin{equation}
  \left<k_T\right>_\mathrm{hadronization}\approx
  1/\sqrt{0.9\log\left(s/s_0\right)}.
  \label{eqn_typical_regge_momentum}
 \end{equation}
  For $s\approx10s_0$, which can occur when, for example, $\sqrt{s_0}=1$
GeV and $\sqrt{s}\approx 3$ GeV, we have a transverse momentum of 0.6
GeV. This gives a measure of the extent of the violation of local
parton-hadron duality due to the colour-singlet phase.

  For lighter dipoles, the transverse momentum becomes greater. However, 
the transverse momentum cannot exceed a half of the dipole mass. From 
eqn.~(\ref{eqn_regge_two_to_two}), we see that the turn-over should 
occur near $s=s_0$.

  A natural choice of $s_0$ is obtained by comparing against the 
Veneziano model. Corresponding to eqn.~(\ref{eqn_regge_two_to_two}), we 
have a Veneziano amplitude:
 \begin{equation}
  A=\beta\frac
    {\Gamma\left(\ell-\alpha(t)\right)\Gamma\left(\ell-\alpha(s)\right)}
    {\Gamma\left(\ell-\alpha(s)-\alpha(t)\right)}.
  \label{eqn_veneziano_two_to_two}
 \end{equation}
  In the Regge limit, by applying the Stirling factorial approximation, 
we recover eqn.~(\ref{eqn_regge_two_to_two}) with the extra constraints 
$\beta(t)=\mathrm{const.}$ and $s_0=1/\alpha'$, where $\alpha'$ is the 
slope. For $\alpha'=0.9$ as in eqn.~(\ref{eqn_traj_roomaf}), we have 
$s_0=1.1$~GeV$^2$. Using the same $\alpha'$ in the $s$- and $t$-channels 
is justified since $\alpha'$ corresponds to the inverse mesonic string 
tension, which is physically, although not necessarily in Regge 
phenomenology, a universal constant.
  $s_0=1.1$~GeV$^2$ implies that the typical transverse momentum 
generated in hadronization is at most $\sim0.5$~GeV, even for the 
lighter dipoles.

  For the sake of comparison, the HERWIG cluster mass cut-off has the 
default value of 3.5 GeV so that the typical cluster mass is about a 
half of this, and so the typical transverse energy is about 0.9 GeV. 
Clusters are the colour-connected units that form the seed of 
hadronization in HERWIG. PYTHIA has the $k_T$ distribution generated 
artificially by a double Gaussian distribution, and the width $\sigma$ 
of the primary Gaussian distribution has the default value of 0.36 GeV. 
This does not imply that hadronization in HERWIG is harder than in 
PYTHIA, as will be demonstrated by a simulation in 
sec.~\ref{sec_montecarlo_threebody}. One of the reasons is that we have 
so far neglected the hadron masses.

  The ratio of the yield of hadron pairs that require heavier flavour 
exchange in the hadronization phase, to the yield of the states that 
only require the exchange of light flavours, is given as a function of 
their invariant mass $M^2_\mathrm{inv}$ as:
 \begin{equation}
  \propto \left(M^2_\mathrm{inv}\right)^{2(\alpha_0-0.5)}.
 \end{equation}
  Here $\alpha_0$ is the intercept of the exchanged trajectory, and is 
less than 0.5. We have squared the amplitude to obtain the probability. 
We have assumed that the effect due to the difference in the slope 
$\alpha'$ of the two trajectories can be ignored.

  A possible practical application of the above formula would be in 
baryon pair production in jets. The largest $\alpha_0$ for baryons is 
0.0, corresponding to one of the $\Delta/N$ trajectories \cite{storrow}. 
The relative baryonic yield is therefore proportional to the inverse of 
the invariant mass squared.

  The physics of three-body baryonic decay of $B$-mesons \cite{haiyang} 
is subject to the same consideration. We can understand the enhancement 
of the three-body decays to two-body decays as the suppression of the 
baryon-exchange when the baryon pair mass is large. This view is 
supported by the so-called `threshold effect', i.e., the tendency that 
the baryon-antibaryon pair is formed with small invariant mass.
  A similar phenomenon is seen in the production of $p\bar p$ in 
low-virtuality $\gamma\gamma$ collision at Belle \cite{bellep}, where 
additional mesons often accompany $p\bar p$, particularly when 
sufficiently above the threshold \cite{wanting}.

 \section{Factorization of the $N$-point amplitude}\label{sec_npoint}
 \label{sec_cascade_amplitude}

 \begin{figure}[ht]{
 \centerline{\epsfig{file=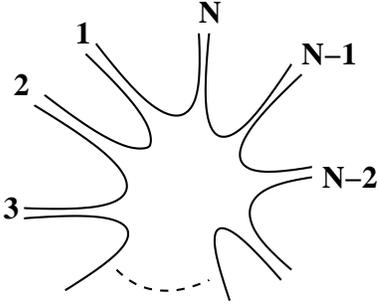, width=5cm}}
 \caption{The $N$-point amplitude.
\label{fig_knnpoint}}}\end{figure}

  The $N$-point Veneziano amplitude corresponds to the scattering of 
open bosonic string shown in fig.~\ref{fig_knnpoint}. This is often 
expressed in a cyclically symmetric form \cite{koba-nielsen}, but for 
our purpose, the formulation of ref.~\cite{bardakciruegg} is more 
convenient. We have:
 \begin{eqnarray}
  A&=&g^{N-2}\prod_{i=2}^{N-2}\int_0^1du_i\
  u_i^{-1-\alpha(t_i)}\ (1-u_i)^{-1-\mu^2}\nonumber\\&\times&
  \prod_{j=2}^i \left(1-\prod_{k=j}^iu_k\right)^{-2p_j\cdot p_{i+1}}.
  \label{eqn_bardakciruegg}
 \end{eqnarray}
  The approximations made in the above equation, in terms of the 
trajectory, is $\alpha(t)=-\mu^2+t$ where $\mu$ is the mass of the 
external legs. This approximation will be removed in the factorized 
formula to be derived later on.
  $g^{N-2}$ is the coupling factor. $t_i$ are the square of the momenta 
that flow in between the $i$'th and the $i+1$'th legs, defined by:
 \begin{equation}
  t_i=\left(\sum_{j=1}^{i}p_j\right)^2
     =\left(\sum_{j=i+1}^{N}p_j\right)^2.
  \label{eqn_definition_of_t}
 \end{equation}

  The multi-Regge limit is obtained in eqn.~(\ref{eqn_bardakciruegg}) by 
the approximation that $u_i$ are small. However, when considering the 
application to hadronization, this multi-Regge formula is not very 
convenient, for two reasons.
  The first is that resonances are not incorporated.
  The second is that it is difficult to construct an iterative evolution 
algorithm based on this equation, that starts from two incoming objects.

\begin{figure}[ht]{
 \centerline{\epsfig{file=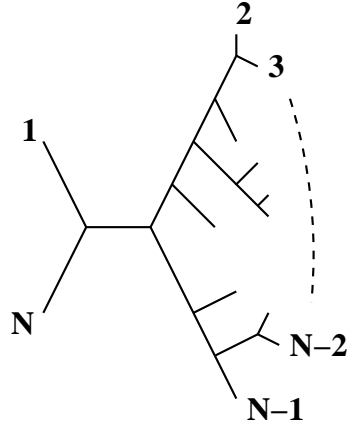, width=4.5cm}}
 \caption{Cascade-decay picture of the $N$-point amplitude.
 \label{fig_cdnpoint}}}\end{figure}

  Let us derive a different limit of eqn.~(\ref{eqn_bardakciruegg}), 
corresponding to the cascade-decay configuration illustrated in 
fig.~\ref{fig_cdnpoint}.

  In eqn.~(\ref{eqn_bardakciruegg}), we consider a resonance in the
$l$'th and $l+1$'th legs. We write:
 \begin{equation}
  2p_l\cdot p_{l+1}+\mu^2=(p_l+p_{l+1})^2-\mu^2
  =\alpha(s_l)=n-\varepsilon.
  \label{eqn_scalar_decay_kinematics}
 \end{equation}
  $n$ is an integer which corresponds to the maximum spin in this 
channel.
  The integral is dominated by the region $u_l\sim1$.
  If $u_{l-1},u_{l+1}<\!\!<1$, we find that:
 \begin{eqnarray}
  &&\lim_{\varepsilon\to+0}A(1,\ldots,l-1,l,l+1,\ldots,N)
  \nonumber\\ &&\approx
  A(1,\ldots,l-1,l+1,\ldots,N) \nonumber\\
  &&\quad\times\
  g\lim_{\varepsilon\to+0}\int_0^1du_l\
  u_l^{-1-\alpha(t_l)}\ (1-u_l)^{-1-n+\varepsilon} \nonumber\\
  &&= A_\mathrm{production}\times A_\mathrm{decay}.
  \label{eqn_bardakciruegg_factorization}
 \end{eqnarray}
  The amplitude thus factorizes into the production part and the decay 
part. This is distinct from the usual resonance factorization property 
of the $N$-point amplitude \cite{koba-nielsen}, since the latter is in 
general not amplitude-squared factorizable and is therefore of limited 
use.

  The decay part is just the Euler beta function. With the usual 
analytization convention, this becomes:
 \begin{equation}
  \lim_{\varepsilon\to\pm0}A_\mathrm{decay}=
  g\frac{\Gamma(-n+\varepsilon)\Gamma(-\alpha(t_l))}
  {\Gamma(-\alpha(t_l)-n+\varepsilon)}.
  \label{eqn_decayamplitude_distribution}
 \end{equation}
  The form of the amplitude is identical with the 4-point Veneziano 
amplitude excepting the difference in the coupling factor.
  If $n$ is large, the angular distribution has Regge behaviour. 
Resonance leads to the factorization in 
eqn.~(\ref{eqn_bardakciruegg_factorization}), but the Regge angular 
distribution is independent of resonance. In general, so long as the 
mass of the decaying system is large, one can consider the dynamics to 
be dominated either by resonances, with $u_l\approx1$, or by poles, with 
$u_l\approx0$.

  We can therefore consider a `Regge cascade' iterative procedure, in 
which we have a series of two-body decays, each of which except the 
final ones exhibiting Regge angular distribution. We can avoid 
intermediate configurations that are manifestly non-Regge by requiring 
that for each decay, $s$ is large and the masses of the two decay 
products are small. This can be achieved by the choice of a sensible 
ordering variable, related to the mass. This will be discussed in 
sec.~\ref{sec_ordering_variable}.

  Let us consider the production amplitude in 
eqn.~(\ref{eqn_bardakciruegg_factorization}). This is as given by 
eqn.~(\ref{eqn_bardakciruegg}), but the mass term needs care.

  We consider a link in the cascade-decay chain that has resonant
daughters. Denoting the decay by $[0\to12]$ as shown in
fig.~\ref{fig_twobodydecay}, the decay distribution is given by:
 \begin{eqnarray}
  &&A_\mathrm{decay}(0\to12)\nonumber\\
  &&=g\lim_{\varepsilon\to+0}\int_0^1du\
  u^{-1-\alpha(t)}\ (1-u)^{-1-\mu^2-2p_1\cdot p_2}.
 \end{eqnarray}
  This is as before. However, eqn.~(\ref{eqn_scalar_decay_kinematics}) 
needs to be modified to:
 \begin{eqnarray}
  2p_1\cdot p_2+\mu^2&=&(p_1+p_2)^2-p^2_1-p^2_2+\mu^2\nonumber\\
  &=&\alpha(p_0^2)-\alpha(p_1^2)-\alpha(p_2^2)\nonumber\\
  &=&n_0-n_1-n_2-\varepsilon.
  \label{eqn_general_decay_kinematics}
 \end{eqnarray}
  In terms of the 4-point Veneziano amplitude, we have the replacement:
 \begin{equation}
  \alpha(s)\to \alpha(s)-\alpha(s_1)-\alpha(s_2),
 \end{equation}
  so that the decay amplitude is modified to:
 \begin{eqnarray}
  A_\mathrm{decay}&\sim&
  g\frac{\Gamma(-\alpha(s))\Gamma(-\alpha(t))}
  {\Gamma(-\alpha(s)-\alpha(t))} \nonumber\\&\longrightarrow&
  g\frac{\Gamma(-\alpha(s)+\alpha(s_1)+\alpha(s_2))
  \Gamma(-\alpha(t))}
  {\Gamma(-\alpha(s)+\alpha(s_1)+\alpha(s_2)-\alpha(t))}.
  \label{eqn_general_decay_amplitude}
 \end{eqnarray}
  The Regge limit of the amplitude is given by:
 \begin{equation}
  A_\mathrm{decay}(0\to12)\sim g\Gamma(-\alpha(t))
  (-\alpha(s)+\alpha(s_1)+\alpha(s_2))^{\alpha(t)}.
  \label{eqn_general_decay_amplitude_regge}
 \end{equation}

  For simplicity, we have omitted other permutations of external 
particles. The introduction of some of these so-called `twisted' terms 
\cite{koba-nielsen} result in the signature factors:
 \begin{equation}
  \left(-\alpha(s)\right)^{\alpha(t)}\longrightarrow
  \left(\alpha(s)\right)^{\alpha(t)}
  \left[1\pm e^{-i\pi\alpha^\pm(t)}\right]
 \end{equation}
  The sign between the two terms, i.e., the signature, depends on the
nature of the trajectory being exchanged.

 \section{The density function} \label{sec_density_function}

  We now turn to developing a practical algorithm based on the 
factorization of eqn.~(\ref{eqn_bardakciruegg_factorization}) and the 
decay amplitude of eqn.~(\ref{eqn_general_decay_amplitude_regge}).
  We first calculate the mass distribution for the daughters 1 and 2 in 
fig.~\ref{fig_twobodydecay} by the optical theorem.

  We temporarily introduce the decay width $\Gamma$ in order to keep 
track of the phase space factors.
  The two-body decay width of an object with mass $\sqrt{s}$, expressed 
in terms of $t$, is given by:
 \begin{equation}
  \frac{d\Gamma(0\to12)}{dt}=\frac{1}{16\pi s^{3/2}}
  \overline{\left|A(0\to12)\right|^2}.
 \end{equation}
  The density function is defined by generalizing this to:
 \begin{equation}
  \frac{d\Gamma(0\to12)}{dtds_1ds_2}=\frac{1}{16\pi s^{3/2}}
  \overline{\left|A(0\to12)\right|^2}
  \rho(s_1)\rho(s_2).
  \label{eqn_density_definition}
 \end{equation}

  We evaluate $\rho(s)$ by the optical theorem. Starting from the total 
decay width, which is in general given by:
 \begin{equation}
  \Gamma = \frac1{2S+1}\frac1{\sqrt{s}}
  \mathrm{Im}\left[A(0\to X\to0)\right],
  \label{eqn_general_decay_width_optical}
 \end{equation}
  we obtain:
 \begin{equation}
  \rho(s_1)=\frac{1}{2\pi}\mathrm{Im}\left[A(1\to X\to 1)\right].
  \label{eqn_density_optical}
 \end{equation}
  The amplitude $A(1\to X\to 1)$ can be estimated as:
 \begin{equation}
  g^2\frac{\Gamma(-\alpha(s_1))\Gamma(-\alpha(0))}
  {\Gamma(-\alpha(s_1)-\alpha(0))}\approx
  g^2\Gamma(-\alpha(0))(-\alpha(s_1))^{\alpha(0)},
 \end{equation}
  so that after substituting $g^2=\alpha'\beta$, we have:
 \begin{equation}
  \rho(s_1)\approx\frac{\alpha'\beta}{2\pi}\Gamma(-\alpha(0))
  (\alpha(s_1))^{\alpha(0)}\sin(\pi\alpha(0)).
 \end{equation}
  Let us absorb the $\Gamma$ function and the phase factor in $\beta$.
  There are several possibilities for estimating this coupling 
coefficient. For instance, an order-of-magnitude estimation can be 
obtained by comparing against the $\pi^+\pi^-$ total cross section. We 
have:
 \begin{eqnarray}
  \sigma^\mathrm{tot}(12\to X)&=&\frac1s\mathrm{Im}A(12\to12)|_{t=0}
  \nonumber\\
  &\approx&\frac{\beta}{s}(\alpha' s)^{\alpha(0)}
  \equiv\alpha'\beta(\alpha's)^{\alpha(0)-1}.
 \end{eqnarray}
  From ref.~\cite{ddln}, using coupling factorization, we have:
 \begin{eqnarray}
  \sigma^\mathrm{tot,\ Reggeon}(\pi^+\pi^-)
  &\approx& \frac{\sigma(\pi^+p)\sigma(\pi^-p)}{\sigma(pp)}\nonumber\\
  &=&\frac{27.56\times 36.02}{56.08}s^{-0.4525}\ (\mathrm{mb}).
 \end{eqnarray}
  Hence:
 \begin{equation}
  \beta\approx 20\, \mathrm{mb\,GeV}^2 \approx 50,
  \label{eqn_beta_estimation}
 \end{equation}
  so that:
 \begin{equation}
  \rho(s_1)\approx\frac{50\alpha'}{2\pi}s_1^{0.5475}\approx7s_1^{0.55}.
  \label{eqn_density_estimation}
 \end{equation}
  $\rho$ is measured in units of GeV$^{-2}$ and $s$ is measured in units 
of GeV$^2$.
  $\rho$, in principle, includes the non-continuum, resonance, 
contribution as well as the other trajectories.

  The density thus obtained does not agree with the one obtained by 
integrating over two-body decays. The contribution to the density 
function from two-body decay is:
 \begin{equation}
  \rho(s)_{0\to12}=\int \frac{dtds_1ds_2}{16\pi^2s}
  \left|A_\mathrm{decay}\right|^2\rho(s_1)\rho(s_2).
  \label{eqn_density_real}
 \end{equation}
  This does not agree with eqn.~(\ref{eqn_density_estimation}).
  This situation is familiar from perturbation theory. The single 
emission of a gluon from a parton gives rise to an infrared-divergent 
contribution to the cross-section. This is resolved by adding together 
the virtual corrections of the same order. The Sudakov form factor gives 
an all-order expression for the probability of emission (or no 
emission), and this effectively sums, up to an infrared cutoff, the 
divergent parts of the real and virtual diagrams together.

  Making the correspondence with the perturbative case, 
eqn.~(\ref{eqn_density_optical}) is the all-order sum. 
Eqn.~(\ref{eqn_density_real}) corresponds to the single-emission cross 
section. The `splitting function' is the ratio of the two:
 \begin{eqnarray}
  P_\mathrm{splitting}(\Phi)d\Phi&=&
  \frac{d\rho(s)_{0\to12}}{\rho(s)}\nonumber\\&=&
  \frac1{\rho(s)}
  \frac{dtds_1ds_2}{16\pi^2s}
  \left|A_\mathrm{decay}\right|^2\rho(s_1)\rho(s_2).
  \label{eqn_the_splitting_function}
 \end{eqnarray}
  $d\Phi$ is the phase space element as before.
  Writing down the Sudakov form factor requires the choice of a suitable 
evolution variable.

 \section{The ordering variable} \label{sec_ordering_variable}

  The factors to be considered in proposing the ordering variable for 
cascade decay are:
 \begin{enumerate}

  \item The algorithm generates all configurations and avoids double 
counting.

  \item The ordering should be physical, i.e., there should be Regge 
behaviour at every vertex except the last decay. Vertexes where the 
daughter masses are comparable with the mass of the mother should be 
avoided where possible.

  \item It should be local, i.e., the decay in one branch of the tree 
should not depend on the decay of the other.

 \end{enumerate}
  In view of the above, it would seem that $t_i$ defined in 
eqn.~(\ref{eqn_bardakciruegg}) may be suitable, as any collection of 
$t_i$ has unique ordering so long as there are no degenerate subsets. 
Thus the first point is satisfied.
  We may further improve this ordering and say that the ordering is in 
each branch of the tree, so that in order to generate the decay 
somewhere in one branch, it is not necessary to look up the values of 
$t_i$ in other parts of the tree. This then satisfies the locality 
condition.
  If we require that $|t_i|$ are ordered in the decreasing order, the 
decay with the largest daughter masses would occur first, so that the 
requirement of Regge behaviour at every vertex would be approximately 
satisfied.

  However, the form of eqn.~(\ref{eqn_the_splitting_function}) suggests 
that this may not be the best choice insofar as the ease of event 
generation is concerned. A better choice would involve the squared 
masses $s_1$ and $s_2$. We therefore add the conditions:
 \begin{enumerate}\addtocounter{enumi}{3}
  \item The ordering variable is expressible in terms of the masses of 
the daughters.

  \item An approximate solution is acceptable if the approximation has a 
physical ground.
 \end{enumerate}
  This suggests the quantity:
 \begin{equation}
  \frac{s_1s_2}{s}\approx |t|_\mathrm{min}.
  \label{eqn_evolution_variable_definition}
 \end{equation}
  Unlike $t_i$, $s_i$ depend on the configuration of the tree, and so 
does, to some extent, the above quantity. This would sometimes lead to 
the violation of the uniqueness of ordering.

  The quantity defined above is similar in form to the transverse 
momentum, so that we also choose to call this quantity $k_T^2$. This is 
proportional to the transverse momentum of the emission that would be 
required for this cascade decay, had the decay been caused by the 
emission of a $q\bar q$ pair.
  We also define rapidity $y$ as:
 \begin{equation}
  y=\frac12\log(s_2/s_1).
 \end{equation}
  From the form of $\rho(s)$, we have:
 \begin{equation}
  \frac{\rho(s_1)\rho(s_2)}{\rho(s)}\approx\rho(k_T^2).
  \label{eqn_density_approximation}
 \end{equation}
  Using this relation, eqn.~(\ref{eqn_the_splitting_function}) 
simplifies to:
 \begin{equation}
  P_\mathrm{splitting}(\Phi)d\Phi=
  \frac{dt}{16\pi^2}dk_T^2dy
  \left|A_\mathrm{decay}\right|^2\rho(k_T^2)
  \label{eqn_the_splitting_function_simple}.
 \end{equation}
  Since $|t|_\mathrm{min}\approx k_T^2$ and $A_\mathrm{decay}$, from 
eqn.~(\ref{eqn_general_decay_amplitude_regge}), only has mild dependence 
on $s_1$ and $s_2$, we conclude that the distribution of the splitting 
function is almost flat in $y$.

  The Sudakov form factor, that is, the probability of no decay in 
between two specified phase space boundaries, is given in general by:
 \begin{equation}
  \Delta(\Phi)=\exp\left[-\int^\Phi_{\Phi_0}
  P_\mathrm{splitting}(\Phi')d\Phi'\right].
  \label{eqn_general_sudakov}
 \end{equation}
  Using the simplified expression of
eqn.~(\ref{eqn_the_splitting_function_simple}) and further imposing the
approximations:
 \begin{equation}
  |t|_\mathrm{min}=k_T^2, \qquad 
  A_\mathrm{decay}=g\Gamma(-\alpha(t))(-\alpha(s))^{\alpha(t)},
  \label{eqn_decay_amplitude_approximation}
 \end{equation}
we obtain the estimation:
 \begin{eqnarray}
  &&\Delta(k_T^2,s)\approx
  \exp\Biggl[-\int^{k_T^2}dQ^2dy\frac{\beta}{16\pi^2}\nonumber\\
  &&\times\frac{\Gamma(-\alpha(-Q^2))^2}{2\log(\alpha(s))}
  (\alpha(s))^{2\alpha(-Q^2)}\rho(Q^2)
  \Biggr].
  \label{eqn_approximate_sudakov}
 \end{eqnarray}
  This formula then gives the approximate cascade-decay evolution.
  From this equation, we see that the dominant dynamical factor is in 
the exponent $(\alpha(s))^{2\alpha}$. The integration over $y$ 
approximately yields $\log{s}$, which cancels against the corresponding 
expression in the denominator. Absorbing the other $Q^2$ dependence in 
$\beta$ and after integration, we obtain:
 \begin{equation}
  \Delta(k_T^2,s)\approx\exp\left[\frac{(\beta/100)}{\log\alpha's}
  (\alpha's)^{2\alpha(-k_T^2)}\right].
  \label{eqn_more_approx_sudakov}
 \end{equation}

  This can be contrasted with the perturbative expression 
\cite{odagirihadronization}:
 \begin{equation}
  \Delta(y)=\exp\left[-\int\frac{dQ^2}{Q^2}dy
  \frac{\alpha_S(Q^2)C_F}{\pi}\right],
  \label{eqn_perturbative_sudakov}
 \end{equation}
  Comparing eqns.~(\ref{eqn_approximate_sudakov}) and 
(\ref{eqn_perturbative_sudakov}), we may obtain an effective and 
non-universal `non-perturbative $\alpha_S$'.

 \section{Algorithm for cascade decay}
 \label{sec_cascade_decay_algorithm}

  The proposed algorithm for generating multi-hadron final states is, in 
outline:
 \begin{enumerate}
  \item Terminate the parton shower, for instance, by means of a 
universal coupling. In the course of this process, some or all of the 
gluons become $q\bar q$.

  \item Form colour-singlet units from colour-connected partons. 
Depending on whether they involve gluons or not, these would 
respectively correspond to kinked strings \cite{pythia} and clusters 
\cite{herwig}.
  Kinked strings may, for instance, be treated dipole-by-dipole.
One first chooses a colour-connected dipole in the kinked string and,
during or after its hadronization, turn to the resolution of the
remaining colour by re-interacting with the other colour-connected
parton(s).

  \item The dipole hadronization proceeds by cascade decay. The 
evolution variable is $k_T^2$, defined by 
eqn.~(\ref{eqn_evolution_variable_definition}). An initial value is set 
for this, at about $s/4$ where $s$ is the mass squared of the dipole.
  The simplified form of the Sudakov form factor is given by 
eqn.~(\ref{eqn_more_approx_sudakov}).

  \item Rapidity is generated either as a flat distribution, or by
using, for instance, eqn.~(\ref{eqn_the_splitting_function_simple}).
The masses of the two daughters are calculated from $k_T^2$ and $y$.
$t$ is generated according to $\left|A_\mathrm{decay}\right|^2$.

  \item For each daughter, the initial value for the evolution variable 
is the $k_T^2$ of the decay that just took place. The $t$-channel 
4-momentum is recorded so that the kinematics for the next decay can be 
evaluated without having to sum over the momenta of other branches.

  \item At some stage, the daughters are identified with physical 
resonances. This can be incorporated into the density function $\rho$.

  \item The process is repeated until every branch either has been 
identified with a physical resonance or its $k_T^2$ has reached zero. 
The latter would correspond to a decay into two ground-state mesons.

 \end{enumerate}

  Although our starting point involved an explicit formulation of the 
multi-particle amplitude, the above procedure is more general and 
depends only on the principles underlying the amplitude, namely duality, 
factorisability and Regge behaviour.
  The Regge amplitudes are generalizable to non-scalar ground-state 
hadrons, and it is a simple matter to include flavour.

 \section{A Regge fragmentation function}
 \label{sec_fragmentation}

 \begin{figure}[ht]{
 \centerline{\epsfig{file=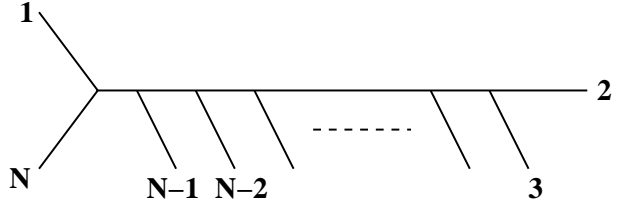, width=8cm}}
 \caption{Fragmentation-function picture of the $N$-point amplitude.
 \label{fig_fragnpoint}}}\end{figure}

  As a special case of the cascade-decay tree configuration, let us 
consider the `fragmentation-function' configuration, shown in 
fig.~\ref{fig_fragnpoint}. This potentially involves cases with very 
heavy daughters so that, for instance, 
eqn.~(\ref{eqn_general_decay_amplitude_regge}) may fail, but we proceed 
by treating it as an approximation. In place of 
eqn.~(\ref{eqn_density_definition}), we have:
 \begin{equation}
  \frac{d\Gamma(0\to12)}{dtds_2}=\frac{1}{16\pi s^{3/2}}
  \overline{\left|A(0\to12)\right|^2}\rho(s_2).
  \label{eqn_density_fragmentation}
 \end{equation}
  The first daughter is the physical, ground state, hadron that splits 
off. We neglect the mass of this hadron.
  The splitting function is:
 \begin{equation}
  P_\mathrm{splitting}(\Phi)d\Phi
  = \frac{\rho(s_2)}{\rho(s)}\frac{ds_2}{s}\frac{dt}{16\pi^2}
  \left|A_\mathrm{decay}\right|^2.
  \label{eqn_fragmentation_splitting_function}
 \end{equation}
  We denote the energy fraction of the branching particle, i.e., 
particle 1, by $z$, so that $s_2=s(1-z)$. Our approximation is valid 
when $z\approx1$. We obtain:
 \begin{equation}
  P_\mathrm{splitting}(\Phi)d\Phi
  = (1-z)^{0.55}dz\frac{g^2dt}{16\pi^2}
  (\alpha(sz))^{2\alpha(t)}.
 \end{equation}
  The evolution variable this time is $z$, so we integrate over $t$. For 
$\alpha(0)=0.55$, we obtain:
 \begin{equation}
  P_\mathrm{splitting}(z)dz
  = dz(1-z)^{0.55}\frac{\beta}{16\pi^2}
  \frac{(\alpha(sz))^{1.1}}{2\log(\alpha(sz))}.
  \label{eqn_splitting_function_an_estimate}
 \end{equation}
  The approximate Regge fragmentation function is:
 \begin{equation}
  f(z)=P(z) \approx z^{1.1}(1-z)^{0.55}\times \mathcal{O}(1).
 \end{equation}

 \section{Signature of gluon splitting in the three-body final state}
 \label{sec_montecarlo_threebody}

  As discussed in the Introduction, the amount of $g\to q\bar q$ forced 
splitting is an ambiguity in our picture of hadronization. In this 
section, we consider the hadronic observables in low-energy 
low-multiplicity events, three-body in particular, that are sensitive to 
this aspect of hadronization. Since HERWIG and PYTHIA represent two 
extreme parametrizations of this splitting, we carry out a simulation 
using these generators.
  Both generators are expected to perform badly with the default 
parameter set, as few-body final states are not what the generators are 
designed for.

  We consider the study of exclusive three-body primary hadron 
production in charm events at BELLE\footnote{We thank A.~Chen for 
suggesting the study of these events.}.
  In both generators, we adopt the default parameter sets. The 
centre-of-mass energy is $10.58$~GeV corresponding to BELLE, and we 
select the charm pair production events, with the matrix element 
correction turned on so that the hard $e^+e^-\to c\bar c g$ 
configuration is generated according to the perturbative matrix element. 
We generate 10000 events in each case.

  We turn off the decay of hadrons in order that the generated hadrons 
are `primary'. The generated final state corresponds to the 
reconstructed few-body events with no further resonances among the final 
state particles.

 \begin{table}[ht]{
  \begin{center}\begin{tabular}{ccc}
  \hline
  Final state & HERWIG & PYTHIA\\
  \hline
  1 hadron & 0 & 0 \\
  2 hadrons & 313 & 269 \\
  3 hadrons & 359 & 1293 \\
  4 hadrons & 2810 & 2795 \\
  5 hadrons & 955 & 3340 \\
  6 hadrons & 3828 & 1699 \\
  7 hadrons & 483 & 494 \\
  8 hadrons & 1150 & 96 \\
  $>8$ hadrons & 102 & 14 \\
  \hline
  Even no.\ hadrons & 8162 & 4860 \\
  Odd no.\ hadrons & 1838 & 5140 \\
  All & 10000 & 10000 \\
  \hline
  \end{tabular}\end{center}
  \label{tab_no_hadrons}
  \caption{The number of primary hadrons, generated using HERWIG and 
PYTHIA.}
 }\end{table}

  We first present a table of the number of the primary hadrons in 
tab.~\ref{tab_no_hadrons}. HERWIG, based primarily on the isotropic 
two-body decay algorithm, normally leads to an even number of hadrons in 
the final state. The exception to the rule occurs when one of the 
clusters is too light to decay into two hadrons. In this case, the 
cluster is identified with the lightest meson with the corresponding 
quantum numbers.
  This predominance of even-numbered multiplicity is considered to be an 
artifact \cite{bryan}, but since this phenomenon is due to the forced 
$g\to q\bar q$ splitting, it is possible that an effect of this nature 
may occur in reality to some extent.

  Both generators predict a typical multiplicity of about four to six 
primary hadrons.

  We now look at the kinematic distribution of the three-hadron final 
state, rejecting the rare events containing baryons. When doing so, we 
propose to make use of a two-dimensional scatter plot in rapidity $y$ 
and the logarithm $\log(k_T)$ of the transverse momentum, both being 
measured against the `jet' direction. We choose this set of observables 
in order that perturbative emissions have an almost flat distribution. 
In the soft limit, we have the DGLAP gluon emission probability 
\cite{preconfinement}:
 \begin{equation}
  dP_\mathrm{emission}=\frac{dQ^2}{Q^2}dy\frac{\alpha_S(Q^2)C}{\pi}.
  \label{eqn_dipole_emission}
 \end{equation}
  $C$ is the colour factor which, for gluon emission from a quark line, 
is $C_F$. We often adopt $Q^2\approx k_T^2$.

  We carry out the simulation at the BELLE centre-of-mass energy of 
$10.58$~GeV. For both the charm quark mass and the charmed meson mass, 
we choose 2 GeV.

  In the three-body case, it is more useful to replace the rapidity and 
$k_T$ by quantities that are more directly measurable. For massless 
particles in the soft limit, for the reaction $[0\to123]$, we write in 
terms of the invariant masses:
 \begin{equation}
  k_T=\frac{M_{13}M_{23}}{M_{123}},
  \qquad
  y=\log\left(M_{23}/M_{13}\right).
  \label{eqn_threebody_y_pt_def}
 \end{equation}
  Particle 3 is soft, and can be identified with the non-charm meson.
  We adopt these definitions even for massive final states. One 
advantage of doing so is that a plot in the $(\log{k_T},y)$ plane is the 
Dalitz plot in the $(M_{13}^2,M_{23}^2)$ plane on the log scale, rotated 
by 45 degrees. The resonances, which form horizontal and vertical lines 
on the Dalitz plot, form straight lines with slope $\pm1$ on this plot.

 \begin{figure}[ht]{
 \centerline{\epsfig{file=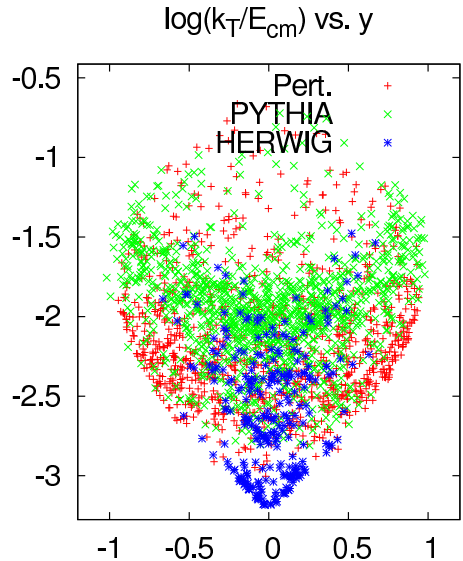,width=6cm}}
 \caption{The $\log(k_T/\sqrt{s})$ vs.\ $y$ scatter plot for charm 
events. We show the results of simulation based on the perturbative 
three-body matrix element with modified $\alpha_S$ (red), PYTHIA (green) 
and HERWIG (blue).
 \label{fig_lkt_y}}}\end{figure}

  In fig.~\ref{fig_lkt_y}, we plot the results of HERWIG and PYTHIA, 
along with the results of simulation based on a perturbative three-body 
matrix element for $e^+e^-\to c\bar cg$ using the modified $\alpha_S$ of 
ref.~\cite{odagirihadronization}.
  The choice of the renormalization scale for the perturbative result is 
$Q^2=k_T^2$. The gluon is identified with the soft particle 3. The peak 
of $\alpha_S$ is at 0.575~GeV.
  The number of generated events is 1000 for the perturbative case and 
10000 for HERWIG and PYTHIA. However, because of the low probability of 
three-body events, the number of accepted events is much less for both 
generators.

  On the vertical axis, a transverse momentum of 1 GeV corresponds to 
$\approx-2.3$, and 0.5~GeV corresponds to $\approx-3$.
  The perturbative distribution is almost flat in rapidity as discussed 
above, with some modification due to the finite mass.

  Turning to the behaviour of the Monte Carlo event generators, we see 
that the behaviour of the two generators is quite distinct.

  The PYTHIA distribution is easier to understand. The band at near 1
GeV, corresponding to $-2$ on the vertical axis, is understood as a
combination of the perturbative gluon emission and the artificially
generated $p_T$, related to the Gaussian width parameters {\tt PARJ(21)
- PARJ(24)}.

  The HERWIG distribution has a complicated structure. The structure 
resembling faint lines at $\pm45$ degrees indicate the typical mass of 
the cluster that decays into one charmed and one non-charmed meson. The 
distribution is in general softer than the PYTHIA distribution because 
the extra particle comes from the decay of a cluster and most of the 
energy in this decay is taken up by the charmed meson. There is little 
remnant of perturbative gluon emission since all of the gluons have 
split into $q\bar q$ and these recombine with the colour-connected 
quark/antiquark to form clusters.

  As mentioned at the beginning of this section, the Monte-Carlo 
predictions should not be trusted for few-body exclusive final states. 
Nevertheless, the experimental investigation of such configurations can 
shed light not only on the areas in which the generators could be 
improved but also on the mechanism of hadronization as a whole. On the 
other hand, there would be considerable and, seemingly \cite{count}, not 
insurmountable difficulty associated with the experimental 
reconstruction of primary hadrons within the plethora of decay products.

 \section{Implementation in Herwig++}\label{sec_implementation}

  We now turn to the implementation of the algorithm developed in the 
previous sections. We choose to adopt Herwig++ \cite{hwpp} as the 
platform. The hadronization algorithm of Herwig++ is based on that of 
HERWIG.

  Our implementation entails:
 \begin{enumerate}

 \item Modified daughter mass distribution in cluster decay, generated 
according to eqn.~(\ref{eqn_more_approx_sudakov}), which can be inverted 
analytically. We impose ordering in $k_T$, and the minimum daughter mass 
is the sum of the constituent quark masses with the corresponding 
flavour. Clusters that are too light to be decayed in the default 
Herwig++ implementation are not decayed.

 \item Option of `cluster rotation', or the Regge smearing of the 
directions of the 3-momenta of the daughter clusters, as well as the 
partons making up the daughter clusters, with respect to the partons 
making up the parent cluster. This has the distribution 
$\propto(\alpha's)^{2\alpha't}$. Here $t$ is the Mandelstam variable for 
the $(2\to2)$ subprocess involving, in the initial state, parent cluster 
partons, and in the final state, the daughter clusters.

 \item Option of pomeron inclusion. By imposing the approximation 
$2\alpha'_{\rm Pomeron}=\alpha'_{\rm Reggeon}$, the resulting Sudakov 
form factor is still analytically invertible.

 \item Option of Regge flavour and baryon generation taking place after 
cluster splitting, with weight $\propto(\alpha's)^{2\alpha(0)}$ and 
adjustable constant multiplicative factor for each flavour and diquark.

 \end{enumerate}
  In eqn.~(\ref{eqn_more_approx_sudakov}), we use the parametrization:
 \begin{equation}
  \beta_{H}\equiv\beta/100=1,
  \quad\alpha'=0.9\ \mathrm{GeV}^{-2},
  \quad\alpha(0)=0.5.
 \end{equation}
  Out of the three parameters, $\alpha'$ and $\alpha(0)$ are fixed by 
Regge phenomenology. The remaining parameter, $\beta_{H}$, is not well 
determined, but the physical results typically depend only on the 
logarithm of $\beta_{H}$. This is because the typical $k_T$ of cluster 
decay is the dominant factor in the decay kinematics.
  The Herwig++ parameter {\tt CLPOW} becomes redundant. We start by 
keeping the values of the other parameters the same as in the default 
set.

  All our simulation results are for $e^+e^-$ interaction at the $Z^0$ 
pole, corresponding to the LEP centre-of-mass energy. We generate 
100,000 events in each case. Our simulation results are compared against 
the default set of experimental data \cite{hwppLEPdata} in Herwig++. 
Both soft and hard matrix-element corrections are switched on.

\begin{figure}[ht]{
 \centerline{\epsfig{file=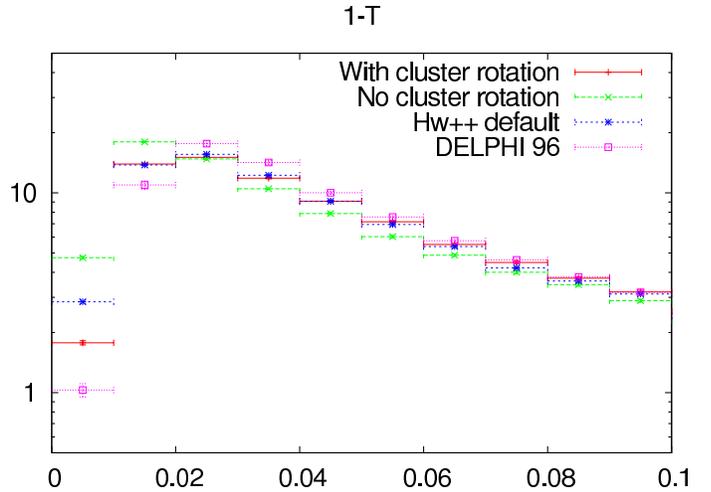,width=9cm}}
 \caption{The distribution of $1-T$, with and without the cluster 
rotation effect.
 \label{fig_thrust}}}\end{figure}

  We first show our simulation result for $1-T$ in 
fig.~\ref{fig_thrust}. After including the cluster-rotation effect, the 
agreement with data is comparable with that of Herwig++, and is better 
in the region of low $1-T$, corresponding to pencil-like events.
  We found similar improvement in most other event-shape observables, 
with the exception of oblateness.
  For spherical events, not shown in the figure, the results are similar 
to those obtained using Herwig++.

\begin{figure}[ht]{
 \centerline{\epsfig{file=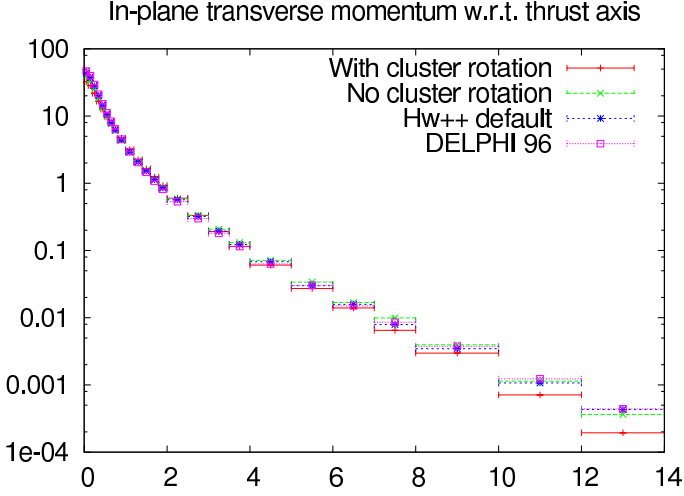,width=9cm}}
 \caption{The distribution of in-plane transverse momentum with respect 
to thrust axis, with and without the cluster rotation effect.
 \label{fig_sp_pT}}}\end{figure}

  In addition to the event shapes, we examined the behaviour of the 
single particle and the identified particle distributions. For the 
former, the results tend to be better than the Herwig++ numbers for 
rapidity, but worse for transverse momenta with respect to thrust and 
sphericity axes. As an example of the latter, fig.~\ref{fig_sp_pT} shows 
the in-plane transverse momentum with respect to thrust axis.
  The transverse momentum distributions improve when 
we omit the cluster rotation, although doing so would ruin the agreement 
with the event shapes. One speculative possibility would be in the 
over-generation of transverse momenta in our combined Regge-Herwig++ 
implementation.

\begin{figure}[ht]{
 \centerline{\epsfig{file=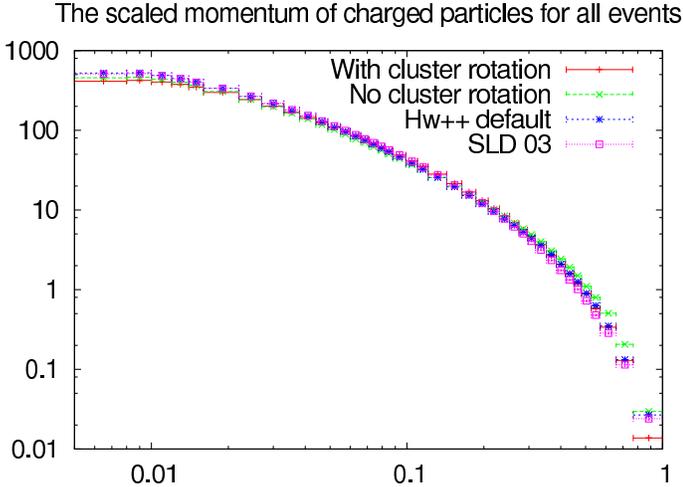,width=9cm}}
 \caption{The distribution of the scaled momentum of charged particles, 
with and without the cluster rotation effect.
 \label{fig_pscaled}}}\end{figure}

  For the identified particle distributions, the results are similar to 
those of Herwig++. The distribution of scaled momentum of charged 
particles, shown in fig.~\ref{fig_pscaled}, shows deficit compared with 
the experimental data both in the very large-$x_p$ and small-$x_p$ 
regions where $x_p$ stands for the scaled momentum. For the large-$x_p$ 
region, we find better agreement when the sample is restricted to 
light-, charm- or bottom-quarks, so that this is possibly not a serious 
deficit.
  In the small-$x_p$ region, the agreement is improved by lowering the 
Herwig++ shower cut-off parameter $\delta$. We found that the flavour 
dependance of the shower cut-off is milder in our approach. Herwig++ 
uses the parametrization:
 \begin{equation}
  Q_g=(\delta-0.3m_q)/2.3,
 \end{equation}
  where $Q_g$ is a gluon virtuality cut-off and $\delta$ is constant,
  but this yields too much radiation from $b$-quarks and too little from 
light quarks in our approach. The choice of $\delta=2.3$ for $b$-quarks 
and $\delta=1.7$ for $c$-quarks was found to be more appropriate.
  An alternative possibility is to introduce the pomeron. We found that 
the substitution:
 \begin{equation}
  (\alpha's)^{2\alpha_\mathrm{Reggeon}}\longrightarrow
  (\alpha's)^{2\alpha_\mathrm{Reggeon}}+
  0.02(\alpha's)^{2\alpha_\mathrm{Pomeron}},
 \end{equation}
  reproduces the observed charged-particle multiplicity. Doing so, 
however, leads to the deterioration in the description of the very 
large-$x_p$ region. On the other hand, this deterioration in the 
large-$x_p$ region makes the agreement with the thrust-like event-shape 
distribution better in the pencil-like region.

\begin{figure}[ht]{
 \centerline{\epsfig{file=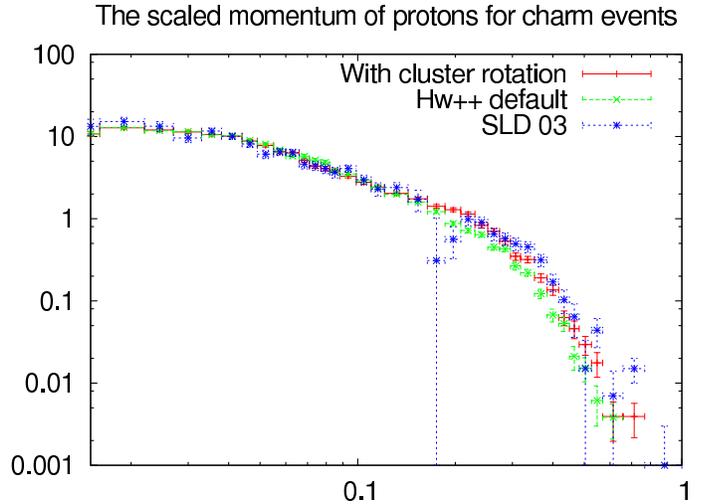,width=9cm}}
 \caption{The distribution of the scaled momentum of protons for charm 
events.
 \label{fig_proton_in_c}}}\end{figure}

  The distribution of the proton was found to be in poor agreement with 
the data and this is similar to the case of Herwig++, but the agreement 
is better in the case of proton production in charm events. This is 
shown in fig.~\ref{fig_proton_in_c}. This seems to imply that we may 
have an incomplete description of either the shower cut-off or 
hadronization in our Regge-Herwig++ amalgamate approach, in the case of 
the light parton from the hard processs.

  For the last item of our implementation, namely that of flavour 
generation during cluster decay, we found no visible effect on any of 
the distributions.

 \section{Conclusions}\label{sec_conclusions}

  We argued that the dynamical conversion of partons to hadrons can be 
effectively factorized into two phases.
  The first is an extended perturbative phase based on an universal
infrared coupling.
  The second is a long-distance hadronization phase mediated by 
colour-singlet resonance--pole dynamics.

  From the consideration of the Veneziano $N$-point amplitude, we argued 
that, essentially because of duality, amplitude-squared factorization is 
applicable to the latter phase as well as the former phase. This 
factorization has the form of cascade decay, where each decay except the 
last one in the chain has Regge angular behaviour.

  Since both phases are factorizable, we can describe the overall 
fragmentation process by merging together the two factorized 
contributions.

  By generalizing the factorization of the $N$-point amplitude, we 
proposed a framework for Monte-Carlo event generation.
  We derived an approximate Regge fragmentation function as a special 
case.

  Our results are applicable generally to multiple hadron production 
from pole--resonance dynamics.

  We made a simple implementation of our hadronization algorithm in 
Herwig++. We found that the event-shapes tend to be better compared with 
the original procedure. On the other hand, the results are mixed in the 
case of single-particle and identified-particle distributions.

 \begin{acknowledgement}

  Acknowledgement.  We thank B.R.~Webber for comments and reading 
through the manuscript, and A.~Chen, C.-H.~Chen, H.-Y.~Cheng, W.-T.~Chen 
and C.-C.~Kuo for informative discussions.


 \end{acknowledgement}

 \end{document}